\documentclass[11pt]{article}
\usepackage{amssymb}
\usepackage{graphicx}
\usepackage{lscape}
\usepackage{theorem}
[1999/03/29 PSNFSS v.7.2
Times font as default roman
: S Rahtz]

\newcommand \be  {\begin{equation}}
\newcommand \bea {\begin{eqnarray}}
\newcommand \ee  {\end{equation}}
\newcommand \eea {\end{eqnarray}}
\newcommand \E {{\rm E}}

\newcommand \w {{\omega}}
\newcommand \Cov {{\rm Cov}}
\newcommand \Var {{\rm Var}}
\newcommand \Dt {{\Delta t}}

\begin{document}

\title{Volatility Fingerprints of Large Shocks:\\
Endogeneous Versus Exogeneous
\footnote{We acknowledge helpful discussions and
exchanges with E. Bacry and V. Pisarenko. This work was partially supported by
the James S. Mc Donnell Foundation 21st century scientist award/studying
complex system.}}
\author{D. Sornette$^{1,2}$, Y. Malevergne$^{1,3}$ and J.-F. Muzy$^{4}$ \\
\\
$^1$ Laboratoire de Physique de la Mati\`ere Condens\'ee CNRS UMR 6622\\
Universit\'e de Nice-Sophia Antipolis, 06108 Nice Cedex 2, France\\
$^2$ Institute of Geophysics and Planetary Physics
and Department of Earth and Space Science\\
University of California, Los Angeles, California 90095, USA\\
$^3$ Institut de Science Financi\`ere et d'Assurances - Universit\'e Lyon I\\
43, Bd du 11 Novembre 1918, 69622 Villeurbanne Cedex, France\\
$^4$ Laboratoire Syst\`emes Physiques de l'Environemment, CNRS UMR 6134 \\
Universit\'e de Corse, Quartier Grossetti, 20250 Corte, France \\
\\
email: sornette@unice.fr, Yannick.Malevergne@unice.fr and muzy@univ-corse.fr\\
fax:  (310) 206 30 51\\
}

\maketitle

\begin{abstract}

Finance is about how the continuous stream of news gets incorporated
into prices. But not all news have the same impact. Can one distinguish
the effects of the
Sept. 11, 2001 attack or of the coup against Gorbachev on
Aug., 19, 1991 from financial crashes such as Oct. 1987 as well as
smaller volatility bursts?
Using a parsimonious autoregressive process
with long-range memory defined on the logarithm of the volatility,
we predict strikingly different response functions of the price volatility
to great external shocks compared to what we term endogeneous shocks, i.e.,
which result from the cooperative accumulation of many small shocks.
These predictions are remarkably well-confirmed
empirically on a hierarchy of volatility shocks.
Our theory allows us to classify two classes of events
(endogeneous and exogeneous)
with specific signatures and characteristic precursors for  the
endogeneous class.
It also explains the origin of endogeneous shocks as the coherent
accumulations of tiny bad news, and thus unify all previous
explanations of large crashes including Oct. 1987.

\end{abstract}

\vskip 1cm

\section{Introduction}

A market crash occurring simultaneously on most of the stock markets
of the world
as witnessed in Oct. 1987 would amount to the quasi-instantaneous
evaporation of trillions
of dollars. Market crashes are the extreme end members of a hierarchy
of market shocks, which shake stock markets repeatedly. Among recent
events still
fresh in memories are the Hong-Kong crash and the turmoil on US
markets on oct. 1997,
the Russian default in Aug. 1998 and the ensuing market turbulence in
western stock markets
and the collapse of the ``new economy'' bubble with the crash of the
Nasdaq index in
March 2000.

In each case,
a lot of work has been carried out to unravel the origin(s) of the crash, so as
to understand its causes and develop possible remedies.
However, no clear cause can usually be singled out. A case in point
is the Oct. 1987 crash, for which
many explanations have been proposed but none has been widely accepted
unambiguously. These proposed causes include
computer trading, derivative securities, illiquidity,
trade and budget deficits, over-inflated prices generated by speculative bubble
during the earlier period, the auction system itself, the presence or absence
of limits on price movements, regulated margin requirements, off-market and
off-hours trading, the presence or
absence of floor brokers, the extent of trading in the cash market
versus the forward
market, the identity of traders (i.e. institutions
such as banks or specialized trading firms), the significance of transaction
taxes, etc. More rigorous and systematic analyses on univariate
associations and multiple
regressions of these various factors
conclude that it is not at all clear what caused the crash \cite{krach87}. The
most precise statement, albeit somewhat self-referencial, is that the most
statistically significant explanatory variable in the October crash can be
ascribed to the normal response of each country's stock market to a worldwide
market motion \cite{krach87}.

In view of the stalemate reached by the approaches attempting to find
a proximal cause
of a market shock, several researchers have looked for more fundamental
origins and have proposed that a crash may be the climax of an
endogeneous instability
associated with the (rational or irrational)
imitative behavior of agents (see for instance
\cite{Orlean4,Orlean95,DSrisk,Shillerexu}).
Are there qualifying signatures of such a mechanism?
According to \cite{DSrisk,Sorquantfin} for which a crash is a stochastic event
associated with the end of a bubble, the detection of such bubble would provide
a fingerprint. A large literature has  emerged on the empirical
detectability of
bubbles in financial data and in particular on rational expectation bubbles
(see \cite{Camerer,AdamSzafarz} for a survey).
Unfortunately, the present evidence
for speculative bubbles is fuzzy and unresolved at best, according to the
standard economic and econometric literature. Other than the
still controversial \cite{Feigenbaum} suggestion
that super-exponential price acceleration \cite{SuperexpJorgen} and
log-periodicity may
qualify a speculative bubble \cite{DSrisk,Sorquantfin}, there are no
unambiguous signatures that would allow one to qualify a market shock
or a crash
as specifically endogeneous.

On the other end, standard economic theory holds that the complex trajectory of
stock market prices is the faithful reflection of the continuous flow of news
that are interpreted and digested by an army of analysts and traders
\cite{Cutler}.
Accordingly, large
shocks should result from really bad surprises. It is a fact that exogeneous
shocks exist, as epitomized by the recent
events of Sept. 11, 2001 and the coup against Gorbachev on Aug., 19, 1991,
and there is no doubt about the existence of
utterly exogeneous bad news that move stock market prices
and create strong bursts of volatility. However,
some could argue that precursory fingerprints of these events were
known to some
elites, suggesting the possibility the action of these informed
agents may have been
reflected in part in stock markets prices.
Even more difficult is the classification (endogeneous versus exogeneous)
of the hierarchy of volatility
bursts that continuously shake stock markets. While it is a common practice to
associate the large market moves and strong bursts of volatility with
external economic, political or natural events \cite{white}, there is not
convincing evidence supporting it.

Here, we provide a clear and novel signature allowing us to distinguish
between an endogeneous and
an exogeneous origin to a volatility shock. Tests on the Oct. 1987 crash, on
a hierarchy of volatility shocks and on a few of the obvious exogeneous shocks
validate the concept. Our theoretical framework combines a rather novel
but really powerful and parsimonious so-called
multifractal random walk with conditional probability calculations.

\section{Long-range memory and distinction between endogeneous and
exogeneous shocks}

While returns do not exhibit discernable correlations beyond a time scale of
a few minutes in liquid arbitraged markets, the historical volatility
(measured as the standard deviation of price returns or more generally
as a positive power of the absolute value of centered price returns)
exhibits a long-range dependence characterized by a
power law decaying two-point correlation function
\cite{Dingetal,Dingranger,A_etal}
approximately following a $(t/T)^{-\nu}$ decay rate with an
exponent $\nu \approx 0.2$.
A variety of  models have been proposed
to account for these long-range correlations
\cite{Grangerding,Baillie,Muller,M_etal,MuzyQF,Muller}.

In addition, not only are returns clustered in bursts of volatility
exhibiting long-range
dependence, but they also exhibit the property of multifractal
scale invariance (or multifractality),
according to which moments $m_q \equiv \langle |r_\tau|^q \rangle$
of the returns at time scale $\tau$
are found to scale as $m_q \propto \tau^{\zeta_q}$,
with the exponent $\zeta_q$
being a non-linear function of the moment order $q$ \cite{Mandel97,M_etal}.

To make quantitative predictions,
we use a flexible and parsimonious model,
the so-called multifractal random walk (MRW)
(see Appendix A and \cite{M_etal,B_etal}), which unifies these two
empirical observations by
deriving naturally the multifractal scale invariance from the volatility
long range dependence.

The long-range nature of the volatility correlation
function can be seen as the direct consequence of
a slow power law decay of the response function $K_{\Delta}(t)$ of the
market volatility measured
a time $t$ after the occurrence of an external perturbation of the volatility
at scale $\Delta t$. We find that the distinct difference between exogeneous
and endogeneous shocks is found in the way the volatility relaxes to its
unconditional average value.

\begin{figure}
\begin{center}
\includegraphics[width=12cm]{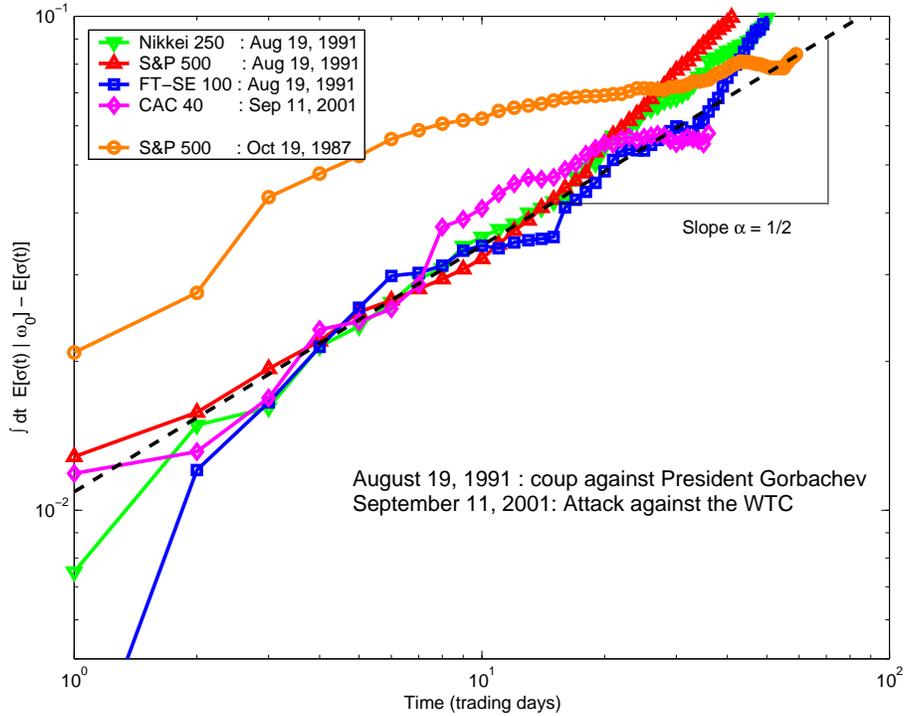}
\end{center}
\caption{\label{fig1}
Cumulative excess volatility at scale $\Delta t=1$ day, that is, 
integral over time of
$\E_{\rm exo}[\sigma^2(t) ~|~ \w_0] -  \overline{\sigma^2 (t)}$, due 
to the volatility
shock induced by the coup against President Gorbachev observed in three
British, Japanese and USA indices and the shock induced by the attack
of September 11, 2001 against the World Trade Center.
The dashed line is
the theoretical prediction obtained by integrating
(\ref{mgmlw}), which gives a $\propto \sqrt{t}$ time-dependence.
The cumulative excess volatility following the crash of October 1987 is
also shown with circles.  Notice that the
slope of the non-constant curve for the October 1987 crash is very 
different from the value $1/2$
expected and observed for exogeneous shocks. This crash and the resulting
volatility relaxation can be interpreted as an endogeneous event.
}
\end{figure}

The prediction of the MRW model (see Appendix B for the technical derivation)
is that the excess volatility $\E_{\rm exo}[\sigma^2(t) ~|~ \w_0] - 
\overline{\sigma^2
(t)}$, at scale $\Delta t$,  due to an
external shock of amplitude $\w_0$ relaxes to zero according to the
universal response
\be
\E_{\rm exo}[\sigma^2(t) ~|~ \w_0] -  \overline{\sigma^2 (t)} \propto
e^{2 K_0 t^{-1/2}}-1 \approx \frac{2K_0}{\sqrt{t}}~,
\label{mgmlw}
\ee
for not too small times,
where  $\overline{\sigma^2 (t)} =\sigma^2 \Dt$ is the unconditional
average volatility.
This prediction is nothing but the response function $K_{\Delta t}(t)$ of the
MRW model to a single piece of very bad news that is sufficient by itself to
move the market significantly. This prediction is well-verified by
the empirical
data shown in figure \ref{fig1}.

On the other hand,  an ``endogeneous'' shock
is the result of the cumulative effect of many small bad news, each
one looking relatively
benign taken alone, but when taken all together collectively along the full
path of news can add up coherently due to the long-range memory of
the volatility
dynamics to create a large ``endogeneous'' shock. This term
``endogeneous'' is thus
not exactly adequate since prices and volatilities are always moved
by external news. The
difference is that an endogeneous shock in the present sense
is the sum of the contribution of many ``small''
news adding up according to a specific most probable trajectory. It
is this set of
small bad news prior to the large shock that not only led to it but
also continues to influence the dynamics of the volatility time
series and creates an
anomalously slow relaxation. Appendix C gives the derivation of the specific
relaxation (\ref{mgngoww}) associated with endogeneous shocks.

\begin{figure}
\begin{center}
\includegraphics[width=12cm]{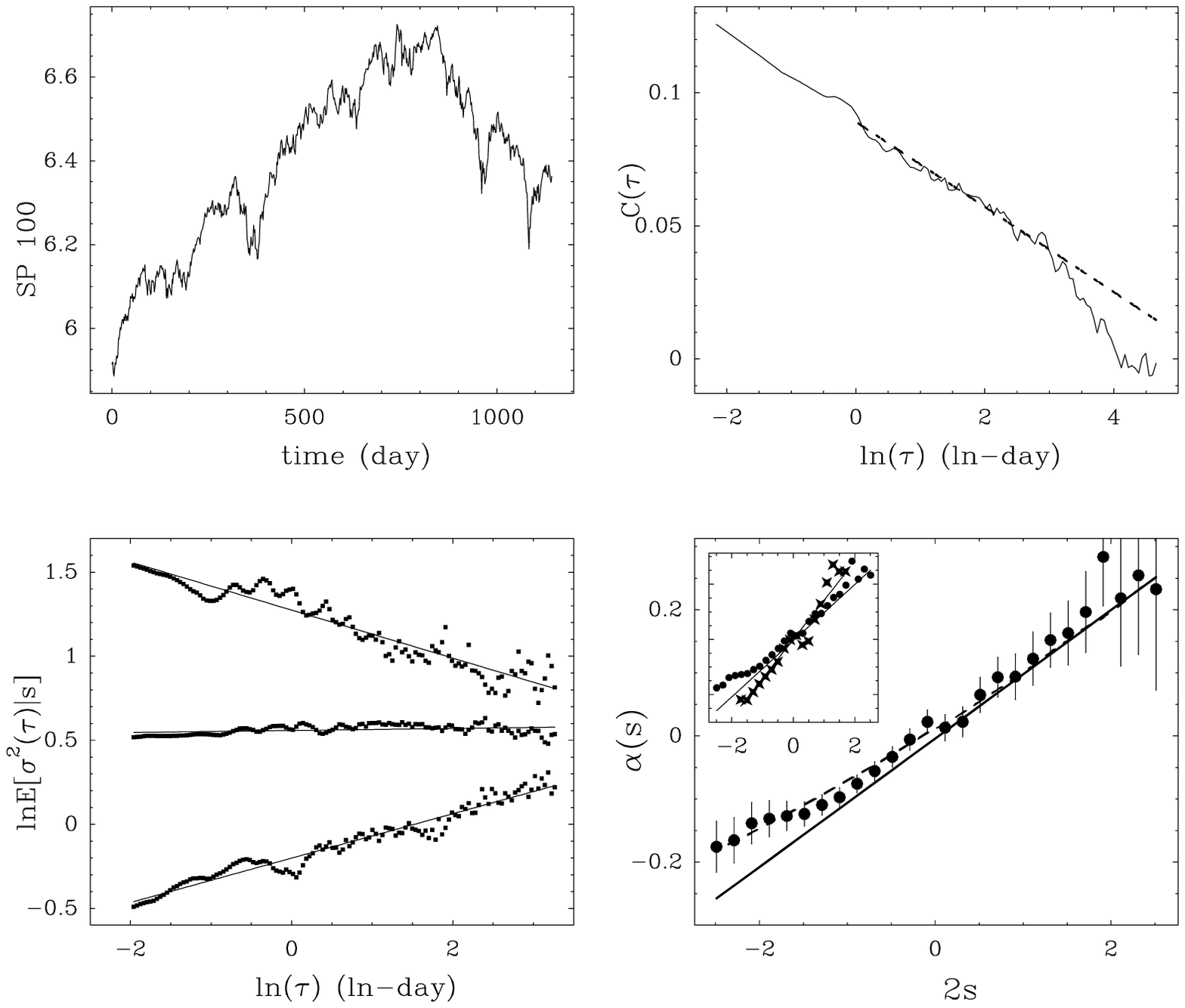}
\end{center}
\caption{\label{fig2} Measuring the conditional volatility response
exponent $\alpha(s)$ for S\&P 100 intradaily time series as a
function of the endogeneous shock amplitude parameterized by $s$, defined
by (\ref{ngwllws}).
(a) The original $5$ minute intradaily time series from 04/08/1997
to 12/24/2001. The $5$ minute de-seasonalized
squared returns are aggregated
in order to estimate the $40$ minutes and daily volatilities.
(b) $40$ minute log-volatility covariance $C_{40}(\tau)$ as a function
of the logarithm of the lag $\tau$. The MRW theoretical curve
with $\lambda^2 = 0.018$ and $T = 1$ year (dashed line) provides
an excellent fit of the data up to lags of one month.
(c) Conditional volatility response $\ln(\E_{\rm endo}[\sigma^2(t) ~|~ s])$
as a function of $\ln(t)$ for three shocks with amplitudes given by $s=-1,0,1$.
(d) Estimated exponent $\alpha(s)$ for $\Dt = 40$ minutes ($\bullet$)
as a function of $s$.
The solid line is the prediction corresponding to
Eq.~(\ref{expoendo}).
The dashed line corresponds to the empirical MRW estimate obtained
by averaging over 500
Monte-Carlo trials. It fits more accurately for negative $s$ (volatility
lower than normal) due to the fact that the estimations of the variance
by aggregation over smaller scales is very noisy for small variance values.
The error bars give the 95 \% confidence intervals estimated
by Monte-Carlo trials of the MRW process.
In the inset, $\alpha(s)$ is compared for $\Dt = 40$ minutes ($\bullet$)
and $\Dt = 1$ day ($\times$).
}
\end{figure}

Figure \ref{fig2} reports empirical estimates
of the conditional volatility relaxation after local maxima
of the S\&P100 intradaily series made of
5 minute close prices during the period from 04/08/1997 to 12/24/2001
(figure 1(a)).
The original intraday squared returns have been
preprocessed in order to remove the U-shaped volatility
modulation associated with the intraday variations of market activity.
Figure \ref{fig2}(b) shows that the MRW model provides a very good fit
of the empirical volatility covariance in a range
of time scales from $5$ minutes to one month.
Fig.~\ref{fig2}(c) plots
in a double logarithmic representation, for the time scale $\Dt = 40$ minutes,
the estimated conditional volatility responses
for $s=1,0,-1$, where the endogeneous shocks are parameterized by
$e^{2s}~\overline{\sigma^2(t)}$. A value $s>0$ (resp. $s<0$) corresponds
to a positive bump (resp. negative dip) of the volatility above
(resp. below) the
average level $\overline{\sigma^2(t)}$. The straight lines are the
predictions (Eqs.~(\ref{endoscaling},\ref{expoendo}))
of the MRW model and qualify power law responses whose
exponents $\alpha(s)$ are continuous function of the shock amplitude $s$.
Figure 1(d) plots the conditional response exponent
$\alpha(s)$ as a function of $s$ for the two time scales
$\Dt = 40$ minutes and $\Dt =$ 1 day (inset).
For $\Dt = 40$ minutes, we observe that $\alpha$ varies between
$-0.2$ for the largest positive shocks to $+0.2$ for the largest
negative shocks, in excellent agreement with MRW estimates
(dashed line) and, for $\alpha \geq 0$, with Eq. (\ref{expoendo})
obtained without any adjustable parameters.
\footnote{The deviation of $\alpha(s)$ from expression (\ref{expoendo})
for negative $s$, originates from the error in the volatility
estimation using a sum of squared returns. The smaller the sum of
squared returns, the larger the error is. As $\Dt$ increases, this error
becomes negligible}
The error bars represent the 95 \% confidence intervals estimated
using 500 trials of synthetic MRW with the same parameters
as observed for the S\&P 100 series.
By comparing $\alpha(s)$ for different $\Dt$ (inset), we can see
the the MRW model is thus able
to recover not only the $s$-dependence of
the exponent $\alpha(s)$ of the conditional
response function to endogeneous shocks but also its time scale $\Dt$
variations:
this exponent increases as one goes from fine to coarse scales.
Similar results are obtained for other intradaily
time series (Nasdaq, FX-rates, etc.). We also obtain the same results
for 17 years of daily return times series of various indices (French, German,
canada, Japan, etc.).

In summary, the most remarkable result is the qualitatively different
functional dependence of the response (\ref{mgmlw}) to an
exogeneous compared to the response (\ref{endoscaling},\ref{expoendo})
to an endogeneous shock. The former gives
a decay of the burst of volatility $\propto 1/t^{1/2}$ compared to
$1/t^{\alpha(s)}$ for endogeneous shocks with amplitude
$e^{2s}~\overline{\sigma^2(t)}$,
with an exponent $\alpha(s)$ being a linear function of $s$.

\section{Discussion}

What is the source of endogeneous shocks characterized by the response function
(\ref{mgngoww})? Appendix D and equation
(\ref{mgnvs}) predict that the expected path of the continuous
information flow prior to the endogeneous shock
grows proportionally to the response function $K(t_c-t)$ measured in
backward time to
the shock occuring at $t_c$.
In other words, conditioned on the observation of a large endogeneous shock,
there is specific set of trajectories of the news flow that led to
it. This specific
flow has an expectation given by (\ref{mgnvs}). This result allows us
to understand
the distinctive features of an endogeneous shock compared to an
external shock. The later
is a single piece of very bad news that is sufficient by itself to
move the market
significantly according to (\ref{mgmlw}). In contrast, an ``endogeneous'' shock
is the result of the cumulative effect of many small bad news, each
one looking relatively
benign taken alone, but when taken all together collectively along the full
path of news can add up coherently due to the long-range memory of
the log-volatility
dynamics to create a large ``endogeneous'' shock. This term
``endogeneous'' is thus
not exactly adequate since prices and volatilities are always moved
by external news. The
difference is that an endogeneous shock in the present sense
is the sum of the contribution of many ``small''
news adding up according to a specific most probable trajectory. It
is this set of small bad news prior to the large shock that not only
led to it but
also continues to influence the dynamics of the volatility time
series and creates the
anomalously slow relaxation (\ref{mgngoww}).

In this respect, this
result allows us to rationalize and unify
the many explanations proposed to account for the Oct. 1987
crash: according to the present theory, each of the explanations is
insufficient to
explain the crash; however, our theory suggests that it is the
cumulative effect of many
such effects that led to the crash. In a sense, the different commentators and
analysts were all right in attributing the origin of the Oct. 1987
crash to many different factors
but they missed the main point that the crash was the extreme response of the
system to the accumulation of many tiny bad news contributions.
To test this idea, we note that the decay of the volatility response
after the Oct. 1987 crash
has been described by a power law $1/t^{0.3}$ \cite{Mantegna}, which is in line
with the prediction of our MRW theory with equation (\ref{expoendo})
for such a large shock (see also figure 2 panel d).
This value of the exponent is still
significantly smaller than $0.5$. Figure 1 demonstrates further the difference
between the relaxation of the volatility after this event shown with
circle and those following the exogenous coup against Gorbachev and 
the September 11 attack.
There is clearly a strong constrast which qualifies
the Oct. 1987 crash as endogeneous,
in the sense of our theory of ``conditional response.''
This provides an independent confirmation of the
concept advanced before in \cite{DSrisk,Sorquantfin}.

It is also interesting to compare the prediction (\ref{mgngoww}) with
those obtained with a linear autoregressive model of the type
(\ref{mbhmle}), in which
$\w(t)$ is replaced by $\sigma(t)$. FIGARCH models fall in
this general class.
It is easy to show in this case that this linear (in volatility)
model predicts the
same exponent for the response of the volatility to endogeneous
shocks, independently
of their magnitude. This prediction is in stark constrast with the prediction
(\ref{mgngoww}) of the log-volatility MRW model.
The later model
is thus strongly validated by our empirical tests.

\section*{Appendix A: The Multifractal Randow Walk (MRW) model}

The multifractal random walk model
is the continuous time limit
of a stochastic volatility model where log-volatility\footnote{The
log-volatilty is the natural quantity used in canonical stochatic volatility
models (see \cite[and references therein]{SNS98}).} correlations decay
logarithmically. It possesses a nice ``stability'' property related to its
scale invariance property: For each time scale $\Dt \leq T$, the returns at
scale $\Dt$, $r_{\Dt}(t) \equiv \ln [p(t)/p(t-\Dt)]$, can be described as a
stochastic volatility model: \be
r_{\Dt}(t) = \epsilon(t) \cdot  \sigma_{\Dt}(t)
= \epsilon(t)
\cdot e^{\w_{\Dt}(t)}~,
\label{remglww}
\ee
where $\epsilon(t)$ is a standardized Gaussian
white noise independent of $\w_{\Dt}(t)$ and
$\w_{\Dt}(t)$ is a nearly Gaussian process with mean
and covariance:
\bea
\label{muo}
\mu_{\Dt} & = & {1 \over 2} \ln(\sigma^2 \Dt)-C_\Dt(0) \\
C_\Dt(\tau) & = & \Cov[\w_{\Dt}(t), \w_{\Dt}(t + \tau)] =
\lambda^2 \ln \left( \frac{T}{|\tau|+e^{-3/2}\Dt} \right)~.
\label{vo}
\eea
$\sigma^2 \Dt$ is the return variance at scale $\Dt$ and
$T$ represents an ``integral'' (correlation)
time scale. Such logarithmic decay of log-volatility covariance
at different time scales has been demonstrated empirically
in \cite{A_etal,M_etal}. Typical values for $T$ and $\lambda^2$ are
respectively $1$ year and $0.02$. According to the MRW model, the
volatility correlation exponent $\nu$ is related to $\lambda^2$ by
$\nu = 4\lambda^2$.

The MRW model can be expressed in a more familiar form, in which the
log-volatility $\omega_\Dt(t)$ obeys an auto-regressive equation whose
solution reads
\be
\omega_\Dt(t) = \mu_\Dt+\int_{-\infty}^t d\tau~ \eta(\tau)~K_\Dt(t-\tau)~,
\label{mbhmle}
\ee
where  $\eta(t)$ denotes a
standardized
Gaussian white noise and
the memory kernel $K_\Dt(\cdot)$ is a causal function, ensuring that the
system is not
anticipative.
The process $\eta(t)$ can be seen as the information flow. Thus
$\w(t)$ represents the response of the market to incoming information up to the
date $t$. At time $t$, the distribution of $\w_\Dt(t)$ is Gaussian with mean
$\mu_\Dt$ and variance
$V_\Dt = \int_0 ^\infty d\tau~
K^2_\Dt(\tau) = \lambda^2 \ln \left( \frac{Te^{3/2}}{\Delta t} \right)$.
Its covariance, which entirely specifies the random process, is given by
\be
C_\Dt(\tau) = \int_0 ^\infty dt~ K_\Dt(t) K_\Dt(t+|\tau|)~.
\label{eq:ker}
\ee
Performing a Fourier tranform, we obtain
$\hat K_\Dt(f)^2 = \hat C_\Dt(f) = 2 \lambda^2~
f^{-1}\left[\int_0^{Tf}{\sin(t) \over
      t} dt+O\left(f\Dt\ln(f\Dt)\right)\right]$,
which shows that for $\tau$ small enough
\be
K_\Dt(\tau) \sim K_0 \sqrt{\frac{\lambda^2 T}{\tau}} ~~~~~
\mbox{for}~~ \Dt << \tau << T~.
\label{mgmlww}
\ee
This slow power law decay (\ref{mgmlww}) of the memory kernel in (\ref{mbhmle})
ensures the long-range dependence and multifractality of the
stochastic volatility
process (\ref{remglww}). Note that equation (\ref{mbhmle}) for the
log-volatility
$\w_\Dt(t)$ takes a form similar
to but simpler than the ARFIMA models usually defined on the (linear)
volatility $\sigma$ \cite{Baillie}.

\section*{Appendix B: Linear response to an external shock}

Let us assume that new major piece of information $\eta(t) = \w_0 ~\delta(t)$
impinges on the market at some time (taken without loss of generality
to be $t=0$, since the system is stationary). $\w_0$ is the amplitude of
the external shock. Then, using the formalism of Appendix A,
the response of the log-volatility
$\w(t)$, $t>0$ to this shock is
\bea
\w(t) &=& \mu+\int_{-\infty}^t d\tau~ [\w_0 \cdot \delta(\tau) + \eta(\tau) ] ~
K(t-\tau)  \\
&=& \mu + \w_0~ K(t) +  \int_{-\infty}^t d\tau ~ \eta(\tau) ~K(t-\tau)~.
\eea
where, for notation convenience, we have omitted the reference to
the scale $\Dt$.
The expected volatility conditional on this incoming major information
is thus
\bea
\E_{\rm exo} [\sigma^2(t) ~|~ \w_0] &=&  \E_{\rm exo} \left[ 
e^{2w(t)} ~|~ \w_0 \right]
    = e^{2 \w_0~ K(t) } ~ \E \left[ e^{2 \int_{-\infty}^t d\tau ~ \eta(\tau)
~K(t-\tau) } \right]~ = \overline{\sigma^2 (t)} e^{2 \w_0~ K(t) } \\
& \simeq & \overline{\sigma^2 (t)} e^{2 K_0 \sqrt{\lambda^2T \over
t}} \; ~\mbox{for}~\;  \Dt << t << T.
\eea
For time $t$ large enough, the volatility relaxes to its
unconditional average value
$\overline{\sigma^2 (t)} =\sigma^2 \Dt$, so that the
excess volatility $\E_{\rm exo}[\sigma^2(t) ~|~ \w_0] - 
\overline{\sigma^2 (t)}$
due to the external shock decays to zero as
\be
\E_{\rm exo}[\sigma^2(t) ~|~ \w_0] -  \overline{\sigma^2 (t)} \propto
\frac{1}{\sqrt{t}}~.
\label{mgmlw2}
\ee
This universal response of the volatility to an external shock (i.e.,
independent of the amplitude $\w_0$ of the shock) is governed by the
time-dependence (\ref{mgmlww}) of the memory kernel $K(t)$. Note also that
the exponent $1/2$ of this power law decay does not depend on the specific
functional form choosen for the volatility. Indeed, we could have
choosen to define the volatily by the expectation of any power of the absolute
returns, the exponent $1/2$ of the power law would have remain the same.

\section*{Appendix C: ``Conditional response'' to an endogeneous shock}

Let us consider the natural evolution of the system, without any
large external shock,
which nevertheless exhibits a large volatility burst $\w(t=0)=\w_0$  at $t=0$.
   From the definition (\ref{remglww}) with (\ref{mbhmle})
and (\ref{eq:ker}),
it is clear that a large ``endogeneous'' shock requires a special set of
realization of the ``small news'' $\{\eta(t)\}$. To quantifies the
response in such case, we can evaluate $\E_{\rm endo}[\sigma^2(t) ~|~
\w_0]=\E_{\rm endo}[e^{2\w(t)}~|~\w_0]$. Since $\w(t)$ is a Gaussian 
process, the new
process $\w(t)$ conditional on $\w_0$ remains Gaussian, so that
\bea
\E_{\rm endo}[\sigma^2(t) ~|~ \w_0] &=& \E_{\rm endo}[e^{2\w(t)}~|~\w_0]\\
&=& \exp \left( 2 \E[w(t) ~|~ \w_0] + 2 \Var[\w(t) ~|~\w_0] \right).
\label{eq:sigma_cond}
\eea

Due to the still Gaussian nature of the condition log-volatility
$\w(t)$, we easily obtain using (\ref{muo}) and (\ref{vo}),
\bea
\E_{\rm endo}[\w(t) ~|~ \w_0] &=& \E[\w(t)] + 
\frac{\Cov[\w(t),\w_0]}{\Var[\w_0]} \cdot
\left( \w_0 - \E[\w_0] \right),\\
&=& (\w_0 - \mu) \cdot \frac{C(t)}{C(0)} + \mu~,
\label{mnllka}
\eea
and
\bea
\Var_{\rm endo}[\w(t) ~|~ \w_0] &=& \Var[\w(t)] -
\frac{\Cov[\w(t),\w_0]^2}{\Var[\w_0]},\\
&=& C(0)\left(1 - \frac{C^2(t)}{C^2(0)}\right)~.
\eea
Let us set:
\be
    e^{2 \w_0}= e^{2s} \overline{\sigma^2 (t)} \Rightarrow \w_0-\mu = s+C(0)
    \label{ngwllws}
\ee
By subsitution in (\ref{eq:sigma_cond}), we obtain thanks to (\ref{muo}) and
(\ref{vo}),
\bea
\E_{\rm endo}[\sigma^2(t) ~|~ \w_0] &=& \overline{\sigma^2 (t)} \exp \left[
2(\w_0 - \mu) \cdot
\frac{C(t)}{C(0)} - 2\frac{C^2(t)}{C(0)} \right],\\
&=& \overline{\sigma^2 (t)} \left({T \over t}\right)^{\alpha(s)+\beta(t)}
\label{mgngoww}
\eea
where
\bea
\label{expoendo}
\alpha(s) & = & { 2 s \over \ln({T e^{3/2}\over \Dt})}~, \\
\beta(t) & = &  2 \lambda^2 {\ln(t/\Dt) \over \ln(T e^{3/2}/\Dt)} ~.
\eea
Within the range $\Dt < t << \Dt e^{|s| \over \lambda^2}$,
$\beta(t) << \alpha(s)$ and Eq. (\ref{mgngoww}) leads to a power-law behavior:
\be
\label{endoscaling}
\E_{\rm endo}[\sigma^2(t) ~|~ \w_0] \sim t^{-\alpha(s)}
\ee
Notice that $|s|\lambda^{-2}$ provides directly the logarithmic
scaling range over which the power-law can be observed. Since $\lambda^{-2}
\sim 20$, this range can quickly extend over the whole time domain
$[\ln(\Dt), \ln(T)]$.

Along the same line, we can also compute the conditional variance
$\Var\left[ \sigma^2(t) ~|~ \w_0 \right]$. After a little algebra,
we get:
\be
\Var_{\rm endo}\left[ \sigma^2(t) ~|~ \w_0 \right] =
\overline{\sigma^2 (t)}^2 \left({T \over t}\right)^{2\alpha(s)+2\beta(t)}
\left( ({T e^{3/2} \over \Dt})^{4 \lambda^2} ({T \over t})^{-2
\beta'(t)} -1 \right)
\ee
where
\be
    \beta'(t) = 2\lambda^2 {\ln(T/t) \over \ln(T e^{3/2}/\Dt)}
\ee
It is thus easy to obtain the estimate:
\be
\sqrt{\Var_{\rm endo}\left[ \sigma^2(t) ~|~ \w_0
      \right]} \leq \E_{\rm endo}[\sigma^2(t) ~|~ \w_0]
\left( ({T \over \Dt})^{2 \beta(t)} -1 \right)^{1/2}
   \lesssim \sqrt{6 \lambda^2 \ln(t e^{3/2}/\Delta t)} ~~\E_{\rm 
endo}[\sigma^2(t)
~|~ \w_0]
\ee
We thus conclude, that, for $s$  large enough (i.e., $\alpha(s)$
large enough):
\be
{ \E_{\rm endo}[\sigma^2(t) ~|~ \w_0]- \overline{\sigma^2 (t)} \over
\sqrt{\Var_{\rm endo}\left[ \sigma^2(t) ~|~ \w_0 \right]}} \gtrsim {1 \over
\sqrt{6 \lambda^2 \ln(t e^{3/2}/\Delta t)}}
\ee
Over the first decade $\Dt \leq t \leq 10 \Dt$,
the deviation of the conditional mean volatility from the
unconditional volatility $\overline{\sigma^2 (t)}$ is greater than
the conditional variance, which ensures the existence of a strong deterministic
component of the conditional response above the stochastic components.

Expressions (\ref{endoscaling},\ref{expoendo}) are our two main predictions.
These equations predict that
the {\it conditional} response function
$\E_{\rm endo}[\sigma^2(t) ~|~ \w_0]$ of the volatility decays
as a power law $\sim 1/t^{\alpha}$ of the time since the endogeneous
shock, with an
exponent $\alpha \approx \left(2\w_0-\ln(\overline{\sigma^2
(t)})\right) \frac{\lambda^2}{C(0)}$
which depends linearly upon the amplitude $\w_0$ of the shock.
Note in particular, that $\alpha$ changes sign: it is positive
for $w_0 > {1 \over 2} \ln(\overline{\sigma^2 (t)})$
and negative otherwise.

\section*{Appendix D: Determination of the sources of endogeneous shocks}

What is the source of endogeneous shocks characterized by the response function
(\ref{mgngoww})?  To answer, let us consider the
process
$W(t) \equiv \int_{-\infty}^t d\tau~\eta(\tau)$, where  $\eta(t)$ is a
standardized Gaussian white noise which captures
the information flow impacting on the volatility, as defined
in (\ref{mbhmle}).
Extending the
property (\ref{mnllka}),
we find that
\be
\E_{\rm endo}[W(t) ~|~ \w_0] = \frac{\Cov[W(t),\w_0]}{\Var[\w_0]} \cdot
\left( \w_0 - \E[\w_0] \right) \propto \left( \w_0 - \E[\w_0] \right)
\int_{-\infty}^t d\tau ~K(-\tau)~.
\label{mgnvs}
\ee
Expression (\ref{mgnvs}) predicts that the expected path of the continuous
information flow prior to the endogeneous shock (i.e., for $t<0$)
grows like $\Delta W(t) = \eta(t) \Delta t \sim K(-t) \Delta t \sim
\Delta t/\sqrt{-t}$ for $t<0$ upon the approach to the time $t=0$ of the large
endogeneous shock. In other words, conditioned on the
observation of a large endogeneous shock,
there is specific set of trajectories of the news flow $\eta(t)$ that led to
it. These conditional news flows have an expectation given by (\ref{mgnvs}).

\end{document}